\documentclass[11pt]{article}
\usepackage[margin=1in]{geometry}
\usepackage{amsmath,amssymb}
\usepackage{graphicx}
\usepackage[colorlinks=true,linkcolor=blue,citecolor=blue,urlcolor=blue]{hyperref}
\usepackage{authblk}

\title{Representation Transitions Reveal Predictive Structure in Complex Systems\\
\large A Trajectory-Level Reconstruction in a Critical System}
\author[1]{Jiaqi Pan}
\affil[1]{Independent Researcher}
\date{}

\begin{document}
\maketitle
\begin{center}
Corresponding author: \texttt{pjqlegion42@outlook.com}
\end{center}

\begin{abstract}
Trajectories in complex systems often contain structure that is not
captured by the population-averaged, scalar statistics traditionally
used to describe them. Here we reconstruct, through independent
numerical re-simulation rather than literature review, a systematic
search across representational forms --- from single scalar statistics,
through two independent discrete multi-feature classification attempts,
to a trajectory-level structural representation and a diagnostic
surrogate test, to a continuous low-dimensional compression --- applied
to the same underlying trajectory ensemble in a critical complex
system. Predictive failure is not explained by insufficient
correlation: one discrete representation's own strongest individual
feature correlates with the target more strongly ($r=0.588$) than the
eventual successful representation's own headline statistic
($r=0.540$), yet fails to separate the ensemble's two most consequential
cases --- while the continuous representation succeeds, robustly,
across eight independent scale-removal folds ($r=0.677$--$0.717$, all
$p<0.001$). The determining factor was not correlation strength but
whether the representation preserved the structure relevant to
prediction. These results indicate that the choice of representation
determines whether predictive structure in complex systems becomes
accessible --- a scientific variable in its own right, not a downstream
analysis choice.
\end{abstract}

\section{Introduction}

Research on complex systems has long relied on scalar observables,
ensemble averages, and order parameters to characterize collective
behavior --- an approach that has been extraordinarily successful
across statistical physics and beyond \cite{anderson1972more}. This
success rests on an assumption that is rarely stated explicitly: that
the structure relevant to a system's future behavior survives whatever
compression is used to summarize it. When that assumption holds, a
richer dataset or a better-fit model closes any remaining gap. When it
does not, no amount of additional data closes it, because the structure
needed for prediction was discarded before any model had the chance to
use it.

This paper is organized around a narrower and more specific question
than ``what additional statistics might help'': which representation of
a trajectory preserves the structure needed for prediction, and which
representation, however reasonable it looks in advance, discards it?
Answering this requires holding the underlying measurements fixed and
varying only the representation built from them --- not proposing a new
statistic, but testing whether the space in which a trajectory is
described determines which aspects of existing structure can be
recovered from it. \textbf{The claim under test throughout this paper is
epistemic, not causal: whether a given representation renders existing
structure observable, not whether the representation brings that
structure into being.}

A first, reasonable assumption is that some scalar summary statistic ---
computed once per trajectory --- captures what distinguishes
trajectories with different downstream behavior. We test this directly.
Six sequential scalar candidates, and two independently designed
multi-feature vectors evaluated through discrete classification, all
fail to recover the structure separating the system's most behaviorally
divergent cases --- despite feature-target correlations that, in one
instance, exceed what the eventual successful representation itself
achieves. This rules out the simplest explanation for the failure: it
is not that too little was measured, and not that the classification
model was too coarse. The failure is specific to how each representation
partitions the space of possible descriptions --- a representation-family
mismatch, not a data or model deficiency.

This paper establishes three things, each independently verified through
numerical re-simulation rather than assumed from report: (1) a
systematic, reproducible reconstruction of this representation search,
spanning scalar, discrete multi-feature, trajectory-level structural,
and continuous compressed representations; (2) a representation
transition (schematically summarized in Supplementary Fig.~S2) --- a
point at which the representational family itself, not merely its
parameters, changes, and previously inaccessible predictive structure
becomes accessible; and (3) that the specific transition responsible is
a shift from discrete classification to continuous compression, robust
across independent validation folds. Together, these results support
treating representation choice itself as a scientific variable in
complex systems research --- worth testing directly, not assumed
settled by whichever representation a field happens to use by
convention.

\begin{figure}[t]
\centering
\includegraphics[width=0.85\textwidth]{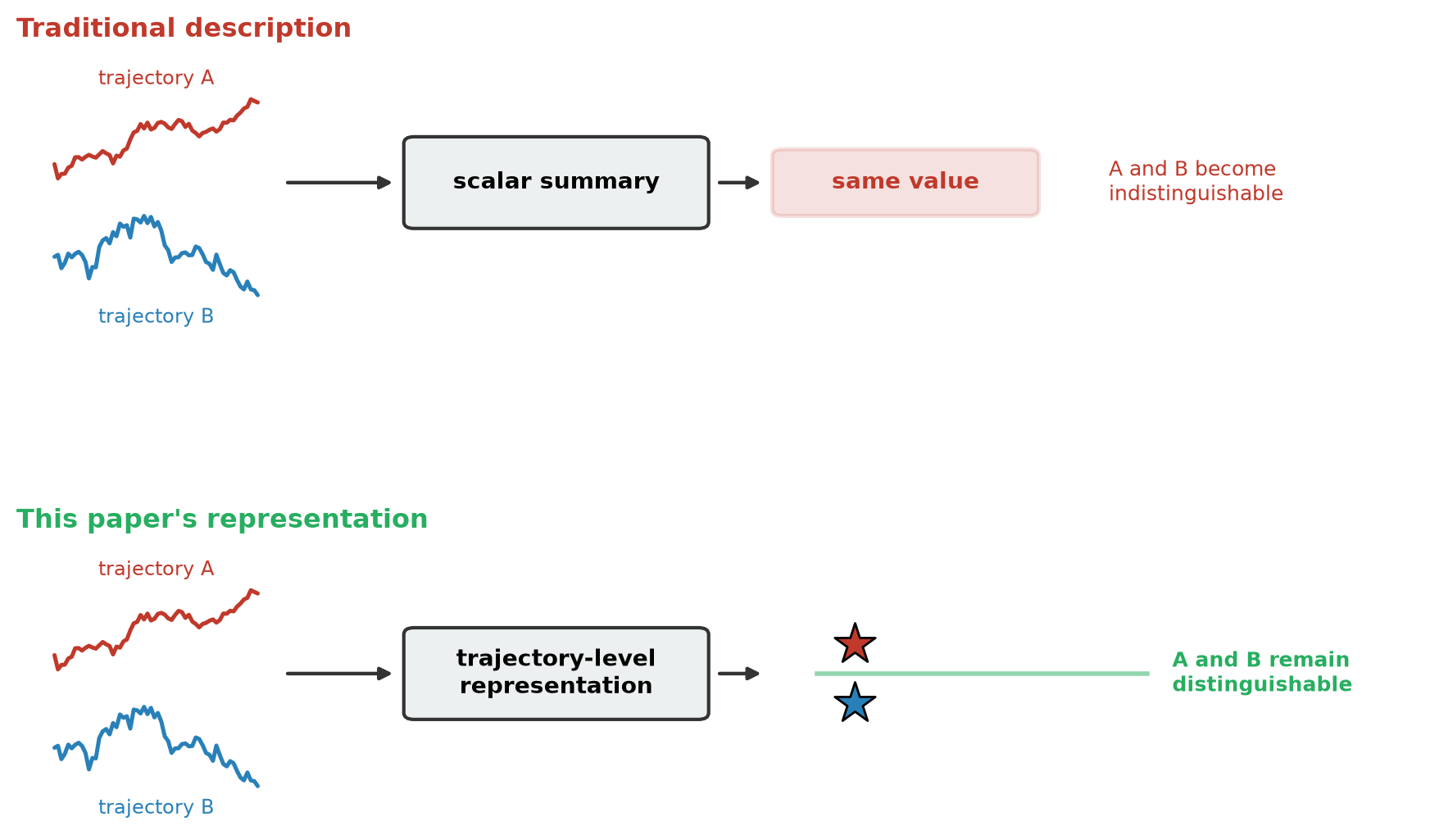}
\caption{Problem definition: same trajectory data, different
representations, different predictive accessibility.}
\label{fig:1}
\end{figure}

Figure~\ref{fig:1} summarizes this problem structure: the same
underlying trajectory data, described under different representations,
yields different predictive accessibility.

\section{Results}

\subsection{Predictive structure exists but is not captured by standard representations}
\label{sec:2.1}

We consider a population of trajectories generated under matched
conditions in a critical complex system, together with a target
variable quantifying each trajectory's own downstream behavioral
divergence (defined in Methods). Before testing any specific
representation, we first establish that this target variable reflects
genuine, reproducible structure --- not measurement noise with no stable
underlying organization: independent re-simulation from the original
random seeds reproduces every reported extreme-trajectory value
exactly, confirming the target is a stable, well-defined property of
the trajectory population, not an artifact of a single realization.

We then test the most direct hypothesis: that some single scalar
statistic, computed once per trajectory, captures what predicts this
target. Six such candidates are tested --- a relaxation-timescale
estimator, a spectral-entropy measure, single-step transition
statistics, higher-order memory measures, and source-coupling strength
--- evaluated across the full trajectory population. None reaches
conventional significance, and in two cases the observed trend runs
opposite to what was hypothesized in advance (Supplementary Appendix
S1, \S2). This establishes only that single-number description is
inadequate --- not that structure is absent, since the target variable
itself remains real and reproducible throughout. \textbf{The signal
exists; whether it is accessible depends on how the trajectory is
represented} --- the question the remainder of this paper tests
directly.

\subsection{Increasing representation complexity does not guarantee predictive success}
\label{sec:2.2}

\begin{figure}[t]
\centering
\includegraphics[width=\textwidth]{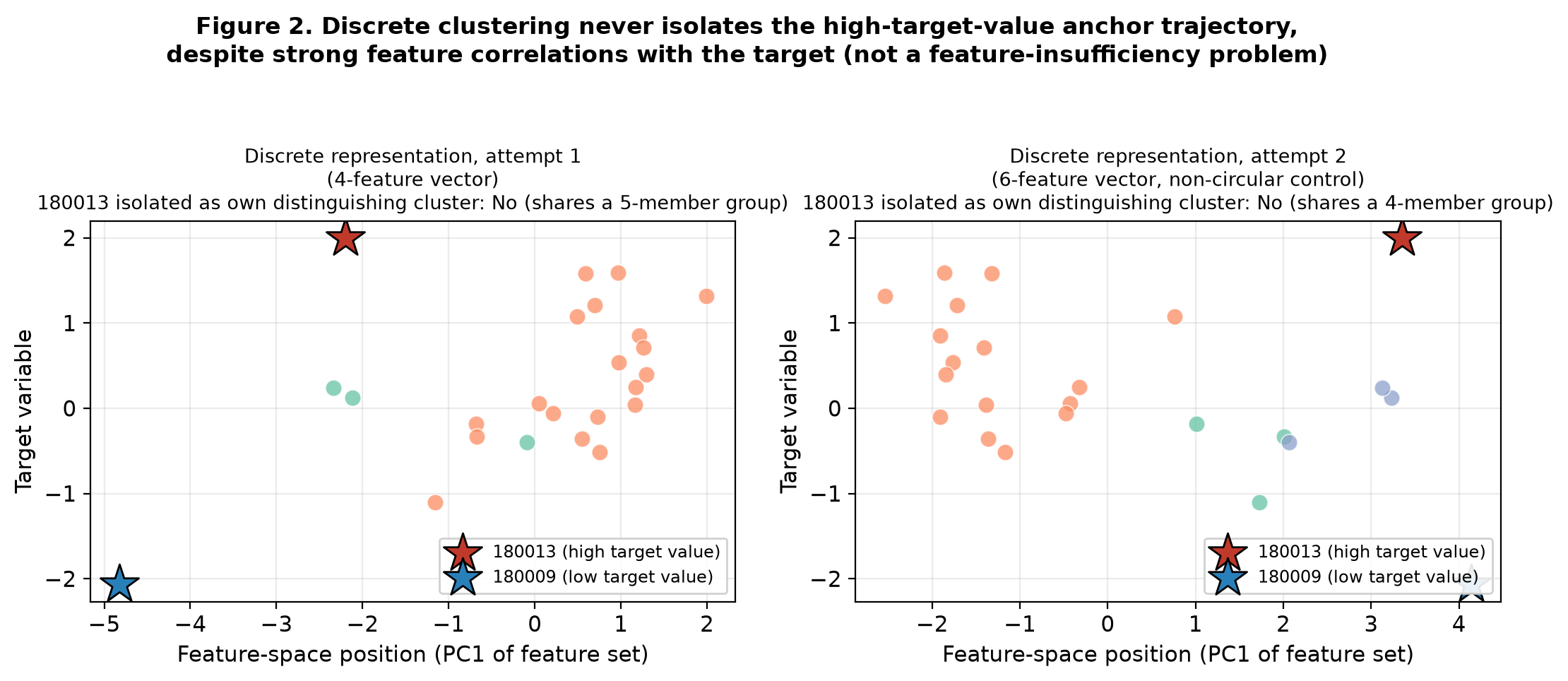}
\caption{Discrete clustering never isolates the high-target-value anchor
trajectory, despite strong feature correlations with the target (not a
feature-insufficiency problem).}
\label{fig:2}
\end{figure}

A natural response to \S\ref{sec:2.1}'s result is to increase
representational richness: describe each trajectory by a vector of
several jointly-computed statistics, and search for discrete structure
--- clusters of behaviorally similar trajectories --- within that
richer feature space. We test this twice, independently: once with a
four-feature vector built from multi-scale measurements of the target
variable's own defining transformation, and once with an independently
designed six-feature vector specifically constructed to exclude
circularity concerns raised by the first attempt. Both recover real,
statistically robust cluster structure (pairwise cluster separation
$p<0.01$ in both cases) --- but in both, the two trajectories with the
most divergent target values are never isolated from the bulk
population: in the first representation they share a 5-member cluster;
in the second, at three clusters, the higher-valued trajectory remains
grouped with three others, never alone (Figure~\ref{fig:2}).

This failure does not reflect weak underlying signal: an individual
feature-target correlation in the first representation reaches
$r=0.588$ --- stronger than the correlation later achieved by the
representation that succeeds (\S\ref{sec:2.3}). We test progressively
richer representations still. A trajectory-level structural
representation built from each trajectory's complete pairwise
transition structure (50 features, evaluated by leave-one-out
regression) fails to predict the target out-of-sample at any tested
regularization strength (best result: $r=0.182$, $p=0.393$). A
diagnostic surrogate test --- generating alternative trajectories that
exactly preserve each original trajectory's own measured pairwise
transition statistics while randomizing everything else --- shows the
trajectories with the most extreme target values lose that extremity
under surrogation in nearly every trial, demonstrating that pairwise
structure, however completely specified, does not contain what makes
these trajectories' values extreme.

Four independent tests of increasing representational complexity
converge on the same result: \textbf{complexity is not adequacy.} Each
attempt adds information to the representation without changing a
shared underlying commitment --- that a trajectory is adequately
described as a finite set of independently-evaluated features,
partitioned or regressed directly. None succeeds, regardless of how
much richer the feature set becomes.

\subsection{Continuous compression reveals trajectory-level predictive structure}
\label{sec:2.3}

Sections \ref{sec:2.1}--\ref{sec:2.2} share an implicit representational
commitment: each represents a trajectory as a finite set of point-valued
features, then reads out a prediction either by direct regression or by
discrete partitioning. We test a different commitment: represent each
trajectory not by a set of point values but by its own continuous
sensitivity profile --- how the trajectory's defining measurement
responds across a graded range of manipulation scales --- and compress
that profile to a single continuous score, without ever partitioning
the population into discrete groups. Principal component analysis
\cite{pearson1901lines,hotelling1933analysis} is used as the simplest
available method for this compression; nothing in what follows depends
on PCA specifically, and this study does not test whether other
continuous, non-partitioning methods would succeed equally well --- only
that discrete partitioning, tested four independent ways in
\S\ref{sec:2.2}, did not, while one continuous alternative did.

\begin{figure}[t]
\centering
\includegraphics[width=\textwidth]{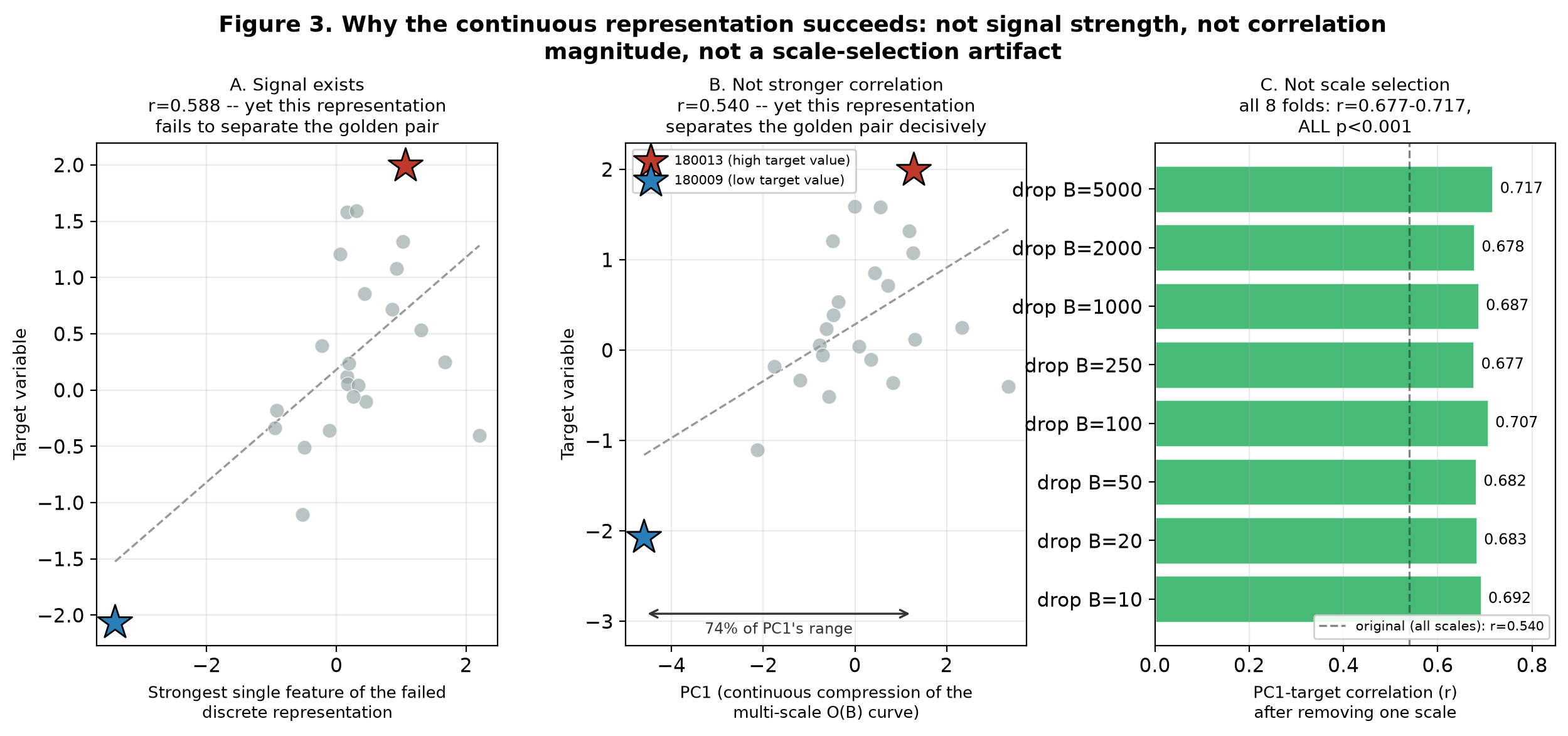}
\caption{Why the continuous representation succeeds: not signal
strength, not correlation magnitude, not a scale-selection artifact.
\textbf{A}: failed representation's signal ($r=0.588$ --- signal
exists). \textbf{B}: successful representation ($r=0.540$, 74\%
separation --- not stronger correlation). \textbf{C}: 8-fold
leave-one-scale-out ($r=0.677$--$0.717$, all $p<0.001$ --- not a
scale-selection artifact).}
\label{fig:3}
\end{figure}

This representation succeeds. The resulting score correlates with the
target variable at $r=0.540$ ($p=0.0065$) --- comparable to the
strongest individual correlations in \S\ref{sec:2.2}'s failed discrete
representations --- but here that correlation translates into actual
separation: the same two trajectories neither \S\ref{sec:2.2} attempt
could isolate span 74\% of this score's full population range, sitting
near opposite extremes (Figure~\ref{fig:3}A--B). A
leave-one-trajectory-out prediction test ($r=0.398$, $p=0.054$) confirms
this is not an in-sample artifact of the fitting procedure.

Because a single correlation could itself be a scale-specific accident,
we test robustness directly: repeating the compression procedure eight
times, each time removing one manipulation scale from the grid before
recomputing the score. The result is stable across every fold
($r=0.677$--$0.717$, all $p<0.001$; Figure~\ref{fig:3}C) --- stronger,
not weaker, than the original estimate, never approaching
non-significance at any fold. The predictive structure this
representation recovers is not an artifact of any single measurement
scale.

The distinguishing property of this representation is not that it
contains more information than \S\ref{sec:2.2}'s discrete attempts ---
by several measures it contains less, being built from projections of
the same underlying measurements onto one axis. The distinguishing
property is that it never requires the population to be partitioned
into discrete groups. \textbf{Predictive structure inaccessible under
every discrete representation tested becomes accessible once the
representation is allowed to remain continuous.}

\subsection{Structural origins beyond scalar observables}
\label{sec:2.4}

Once a representation exists that makes the target variable's
population-level structure accessible, a narrower question becomes
tractable: what specifically distinguishes individual trajectories at
the extremes this representation identifies? We investigate possible
structural origins for two such trajectories, without claiming to
resolve a complete mechanism for either (Supplementary Fig.~S1).

For the trajectory with the highest target value, a direct intervention
test --- combining a reordering of the trajectory's own coarse-grained
segments with a local content-level perturbation --- shows the two
perturbation types interact: applying both together degrades this
trajectory's distinguishing property substantially more than the sum of
their separate effects predicts, replicated under an independent
repeat. This indicates the property depends on a joint structure ---
segment order and content together --- not on either alone.

For the trajectory with the lowest target value, we test and exclude
three candidate structural origins: a second-order sequential
dependency among coarse-grained segment types (excluded via
bias-corrected estimation and bootstrap confidence intervals); an
asymmetry in the trajectory's own driving signal (inconclusive --- this
asymmetry is statistically indistinguishable from a comparison
trajectory with the opposite-sign target value); and whether this
trajectory's own measured properties are simply an extreme, rare draw
relative to a reference ensemble (excluded --- statistically
unremarkable relative to that ensemble). A subsequent population-scale
test ($n=24$, not the original 3-trajectory comparison) further shows
the first trajectory's own apparent distinguishing properties do not
generalize into a population-wide predictive rule; only one of four
tested structural features remains a genuine population-level outlier
at this larger scale.

These results do not establish a general mechanism. They establish that
once the representation-level obstacle identified in
\S\ref{sec:2.1}--\ref{sec:2.3} is removed, specific structural questions
about individual trajectories become directly testable --- some
yielding confirmed, replicated findings; others yielding confirmed
exclusions; none yet composing into a population-general account. This
is reported as the actual outcome of investigating structural origins
once a working representation exists, not as a shortfall against an
unclaimed goal.

\section{Discussion}

This study demonstrates, in one complex system, that whether predictive
organization in a trajectory ensemble becomes accessible depends on the
representation used to describe it. Six scalar candidates and two
independently designed discrete multi-feature representations fail to
make this organization accessible despite genuine underlying signal; a
continuous compression of the same underlying measurements succeeds,
robustly, across independent validation. The determining factor was not
the amount of information in the representation but whether that
representation preserved the structure relevant to prediction.

These results are demonstrated in a single system, using one specific
trajectory ensemble and one specific target variable; whether they
generalize to other systems, other target variables, or other
trajectory ensembles within the same system is not tested here. We do
not claim continuous representations are generally preferable to
discrete ones --- many well-posed prediction problems have genuine
discrete structure, and the result reported here is specific to this
population and this target, not a general ranking of representational
forms. The result should not be interpreted as evidence that principal
component analysis is universally optimal, but as evidence that
changing representational form can expose predictive structure
unavailable in previous descriptions. This study does not propose a new
order parameter, nor any new physical quantity; we study the
accessibility of structure under different representations, rather
than proposing a new physical observable --- the continuous score used
here is a projection of already-standard multi-scale measurements, and
its role in this study is representational, not as a candidate
observable in its own right. Nor do these results argue against
conventional order-parameter or scalar-observable approaches, which
remain successful for the large majority of questions they were
designed to answer; this study instead identifies a specific condition
--- where predictive structure depends on continuous rather than
discrete organization --- under which such approaches can fail even when
the underlying signal is genuinely present.

These observations motivate a broader question: whether representation
selection should be treated as an explicit, testable variable in
complex-systems analysis generally, alongside model choice and data
collection. Answering that question for any system beyond the one
studied here requires an independent test in that system, with its own
falsifiable prediction --- a direction this study motivates but does not
attempt to establish.

\section{Methods}

\subsection*{System and target variable}

The trajectory ensemble is generated under matched conditions in a
critical complex system (full simulation parameters in Supplementary
Table~S1). The target variable quantifies each trajectory's own
downstream behavioral divergence under a defined manipulation (full
definition in Supplementary Table~S2). Every reported value for this
target variable was independently re-derived from the original random
seeds in this study, not taken from prior report.

\subsection*{Why these representations, in this order: search-space fairness}

The representations tested here are not a curated comparison set
assembled after the outcome was known. They correspond to the
representational families that were tried, in sequence, before the
successful representation was found --- each motivated directly by the
specific way the previous one failed:

\begin{enumerate}
\item \textbf{Scalar candidates} (\S\ref{sec:2.1}) were the first tested
  representational family, the natural first hypothesis for any
  single-target prediction problem.
\item \textbf{Discrete multi-feature classification, attempt 1}
  (\S\ref{sec:2.2}) followed directly from scalar failure.
\item \textbf{Discrete multi-feature classification, attempt 2} was
  designed specifically to test whether attempt 1's failure was a
  circularity artifact --- a control, not an arbitrary repeat.
\item \textbf{Trajectory-level structural representation} followed from
  a specific, stated hypothesis: that the feature vectors used so far
  were too compressed.
\item \textbf{The diagnostic surrogate test} followed directly from
  representation 4's own failure, designed to distinguish insufficient
  information from wrong level of representation.
\item \textbf{Continuous compression} (\S\ref{sec:2.3}) is the only
  representation in this sequence that changes the read-out structure
  itself, motivated directly by representation 5's own diagnostic
  result.
\end{enumerate}

Each representation was tested once, evaluated by its own pre-specified
criterion, before the next was designed. Where independent repeats
exist (noted below), they test reproducibility of an already-obtained
result, not a search for a better outcome.

\subsection*{Why the successful representation is not cherry-picked}

Representation 6 is not one of several continuous variants tried until
one succeeded --- it is the only representation in the entire six-step
sequence that changes the representational family itself.
Representations 1--5 all share one commitment (a trajectory reduced to
a finite set of point-valued features, read out by direct regression or
discrete partitioning); representation 6 is the single point in the
sequence where that commitment is dropped. There was exactly one
attempt at this representational family, and it succeeded on that
attempt, evaluated by the same two pre-specified criteria used
throughout.

\subsection*{Reproduction standard}

\begin{itemize}
\item \textbf{Seeds}: every trajectory is independently re-simulated
  from its original random seed for every representation reported.
\item \textbf{Independent reruns}: findings reported as ``confirmed'' or
  ``replicated'' (\S\ref{sec:2.4}) were re-run with a new random seed
  offset and an identical pipeline, reported alongside the first
  result.
\item \textbf{Scale removal}: the robustness result in \S\ref{sec:2.3}
  comes from 8 independent folds, each removing one manipulation scale
  before recomputing the score.
\item \textbf{Statistics}: correlation significance is reported via
  standard Pearson tests throughout; group-difference tests
  (\S\ref{sec:2.4}) use bias-corrected estimation
  (Miller--Madow~\cite{miller1955note}) plus moving-block
  bootstrap~\cite{kunsch1989jackknife} confidence intervals where
  small-sample estimation bias is a known risk (full statistics:
  Supplementary Material).
\end{itemize}

\section*{Data Availability Statement}

All code, simulation parameters, and cached numerical outputs (raw
trajectories and processed results) underlying every figure and
statistic in this manuscript and its Supplementary Material are
publicly archived. The repository is available on GitHub at
\url{https://github.com/863632670/paper-a-reproducibility-public} and
permanently archived on Zenodo under concept DOI
\url{https://doi.org/10.5281/zenodo.22047164}, which resolves to the
latest archived version; a specific version DOI is cited alongside
each archived release for exact reproducibility. No data are excluded
from this archive.

\section*{Acknowledgments}

The author thanks friends and family for their support during this
work. The author also acknowledges the contributions of the many prior
researchers whose work in complex systems, representation learning, and
statistical physics this study builds upon.


\begin{thebibliography}{8}

\bibitem{anderson1972more}
Anderson PW. More Is Different. Science. 1972;177(4047):393--396.
doi:10.1126/science.177.4047.393

\bibitem{metropolis1953equation}
Metropolis N, Rosenbluth AW, Rosenbluth MN, Teller AH, Teller E.
Equation of State Calculations by Fast Computing Machines. J Chem Phys.
1953;21(6):1087--1092. doi:10.1063/1.1699114

\bibitem{ising1925beitrag}
Ising E. Beitrag zur Theorie des Ferromagnetismus. Z Phys.
1925;31(1):253--258. doi:10.1007/BF02980577

\bibitem{schreiber2000measuring}
Schreiber T. Measuring Information Transfer. Phys Rev Lett.
2000;85(2):461--464. doi:10.1103/PhysRevLett.85.461

\bibitem{pearson1901lines}
Pearson K. On Lines and Planes of Closest Fit to Systems of Points in
Space. Philos Mag Ser 6. 1901;2(11):559--572.
doi:10.1080/14786440109462720

\bibitem{hotelling1933analysis}
Hotelling H. Analysis of a Complex of Statistical Variables into
Principal Components. J Educ Psychol. 1933;24(6):417--441.
doi:10.1037/h0071325

\bibitem{miller1955note}
Miller GA. Note on the Bias of Information Estimates. In: Quastler H,
editor. Information Theory in Psychology: Problems and Methods II-B.
Glencoe (IL): Free Press; 1955. p. 95--100.

\bibitem{kunsch1989jackknife}
K\"unsch HR. The Jackknife and the Bootstrap for General Stationary
Observations. Ann Stat. 1989;17(3):1217--1241.
doi:10.1214/aos/1176347265

\end{thebibliography}
\end{document}